\documentclass[aps,prl,twocolumn,superscriptaddress,showpacs,floatfix,amsmath,amssymb]{revtex4-1}
\usepackage{color,graphicx,array,tabularx,bm,url,subcaption}

\begin{document}

\title{Giant vortex dynamics in confined active turbulence}

\author{L. Puggioni}
\affiliation{Dipartimento di Fisica and INFN, Universit\`a degli Studi di Torino, via P. Giuria 1, 10125 Torino, Italy.}

\author{G. Boffetta}
\affiliation{Dipartimento di Fisica and INFN, Universit\`a degli Studi di Torino, via P. Giuria 1, 10125 Torino, Italy.}

\author{S. Musacchio}
\thanks{Corresponding author}
\email{stefano.musacchio@unito.it}
\affiliation{Dipartimento di Fisica and INFN, Universit\`a degli Studi di Torino, via P. Giuria 1, 10125 Torino, Italy.}

\begin{abstract}
We report the numerical evidence of a new state of active turbulence in 
confined domains. By means of extensive numerical simulations of the 
Toner-Tu-Swift-Hohenberg model for dense bacterial suspensions
in circular geometry, we discover the formation a stable, ordered state 
in which the angular momentum symmetry is broken.
This is achieved by self-organization of a turbulent-like flow into a single, giant vortex of the size of the domain.
The giant vortex is surrounded by an annular region close to the boundary,
characterized by small-scale, radial vorticity streaks. 
The average radial velocity profile of the vortex is found to be in 
agreement with a simple analytical prediction.
We also provide an estimate of the temporal and spatial scales of a 
suitable experimental setup comparable with our numerical findings.
\end{abstract}

\date{\today}

\maketitle

Flowing active matter is one of the most fascinating examples of
out-of-equilibrium systems which sits at the intersection between
statistical physics, biophysics and fluid
dynamics~\cite{marchetti2013hydrodynamics,ramaswamy2010mechanics,alert2021active}.
In dense active systems, such as suspensions of bacteria, the collective
motion of the individual swimmers produces complex flows 
at scales much larger than the single swimmer~\cite{sokolov2012physical,wioland2013confinement},
often with chaotic dynamics on several length scales \cite{dombrowski2004self,sokolov2009enhanced,creppy2015turbulence,peruani2012collective,liu2021density}.
In these conditions, the flow produced by the swimmers has several 
similarities with usual, high Reynolds number turbulence, including
the presence of 
coherent structures~\cite{sokolov2007concentration,wensink2012meso,petroff2015fast,nishiguchi2017long},
a wide range of active scales and
anomalous transport~\cite{morozov2014enhanced,ariel2015swarming}
and it leads to states called active turbulence \cite{alert2021active}.

In order to understand and rationalize the experimental observations,
a considerable theoretical effort has been devoted to develop continuous,
coarse-grained descriptions of dense active suspensions 
\cite{simha2002hydrodynamic,saintillan2008instabilities,baskaran2009statistical,peshkov2012nonlinear}. 
More recently, simple models with a reduced number of parameters 
have been introduced
\cite{wensink2012meso,dunkel2013minimal,slomka2015generalized,zhou2014living,bar2020self,shaebani2020computational}, and compared with experimental results 
\cite{wensink2012meso,dunkel2013fluid,ariel2018collective,reinken2020organizing}. 
These minimal models reproduce several 
features of active turbulence such as 
spontaneous flow~\cite{bonelli2016spontaneous,giordano2021activity} and
multiscale dynamics 
\cite{bratanov2015new,james2018vortex,linkmann2019phase,carenza2020multiscale},
clustering~\cite{worlitzer2021motility,worlitzer2021turbulence}
and anomalous diffusion~\cite{mukherjee2021anomalous}.

The numerical studies of these models are often performed in 
two-dimensions.  
This is motivated by the fact that most of the experiments
of bacterial suspensions are conducted in quasi-two-dimensional domains.
Moreover, periodic boundary conditions are often assumed since 
one is interested in the bulk properties of the active flow.
Nonetheless, experiments have shown that confining boundaries can play 
an important role in the organization of the flow 
\cite{wioland2013confinement,lushi2014fluid,wioland2016directed}.
In particular, recent experimental studies have shown that
confining the bacterial suspension in circular micro-wells
induces the formation of a rectified vortex
\cite{wioland2013confinement,lushi2014fluid,beppu2017geometry}.

Here we pursue the investigation of the importance of boundaries
by presenting the results of extensive numerical simulations of the
Toner-Tu-Swift-Hohenberg (TTSH) model~\cite{wensink2012meso,dunkel2013fluid}
confined in two-dimensional circular domains.
We show that the geometrical confinement induces 
the transition to a novel regime, characterized
by the formation of a giant vortex
surrounded by an annular region of elongated vorticity structures 
(streaks). 
By an exploration of the parameter space we find that the appearance
of the giant vortex is a robust feature of the model in the 
presence of confinement, and it occurs in a range of
physical parameters accessible to experiments of bacterial turbulence.

The equation for the coarse-grained collective velocity
field ${\bm u}$ in the TTSH model takes the form 
\begin{equation}
  \partial_t {\bm u} + \lambda {\bm u} \cdot {\bm \nabla} {\bm u}
  = - {\bm \nabla} p - (\alpha + \beta |{\bm u}|^2
  + \Gamma_2 \nabla^2 + \Gamma_4 \nabla^4) {\bm u}\;.
 \label{eq:1}
\end{equation}
The pressure gradient ${\bm \nabla} p$ ensures the incompressibility of 
the flow, ${\bm \nabla } \cdot {\bm u}=0$, 
which is valid for dense suspensions. 
The parameters $\lambda,\alpha,\beta,\Gamma_2,\Gamma_4$
are determined by the properties of the microswimmers.
For pusher swimmers one has $\lambda > 1$, while $\lambda < 1$ 
corresponds to pullers~\cite{alert2021active}.
The Swift-Hohenberg operator $\Gamma_2 \nabla^2 +  \Gamma_4 \nabla^4$ 
selects the characteristic scale $\Lambda = 2\pi \sqrt{2\Gamma_4/\Gamma_2}$
at which the flow is forced by the microscopic
motion~\cite{swift1977hydrodynamic}.
For $\alpha < 0$, the Landau force $(\alpha + \beta|{\bm u}|^2){\bm u}$
promotes the formation of collective motion with velocity 
$U=\sqrt{-\alpha/\beta}$.
Larger values of $|\alpha|$ correspond to stronger aligning
interactions between the swimmers, as in the original Toner-Tu (TT) 
model for flocking~\cite{toner1995long,toner2005hydrodynamics}.
Generalizations of the TTSH model to include coupling with a fluid 
velocity field and compressible flows have been proposed 
\cite{heidenreich2016hydrodynamic,reinken2018derivation,worlitzer2021motility}.

We performed a set of numerical simulations of the TTSH model
confined in two-dimensional circular domains of radius $R$.
No-slip boundary conditions are imposed at the border
of the circular domain by means of the 
penalization method~\cite{angot1999penalization}, by adding the term 
$-\frac{1}{\tau}\mathcal{M}(r){\bm u}$ to the r.h.s. of (\ref{eq:1})
where $\tau$ represents the permeability time and is the smallest dynamical
time in the system. 
The mask function $\mathcal{M}(r) = (\tanh((r-R)/(2 \Delta x))+1)/2$
imposes a sharp decay of the fields at the boundary on a scale of few
grid points with spacing $\Delta x$. 
Numerical integration of (\ref{eq:1}) supplemented with the penalization term
is obtained by fully dealiased pseudospectral code with fourth-order 
Runge Kutta time scheme.
The parameters of the simulations are 
$\Gamma_2=2$, $\Gamma_4=1$, $\beta=0.01$, $\lambda=3.5$, $\tau =10^{-3}$ 
and we vary $\alpha$ on the values $(-2.0;-1.75;-1.5;-1.25)$ and 
$R/\Lambda$ on $(16; 23; 31)$.
For the analysis, we decompose the velocity field in the radial 
and angular components
${\bm u} = u_r \hat{\bm r} + u_\varphi \hat{\bm \varphi}$
which define the radial and angular kinetic energies
$E_r = \frac{1}{2}\langle u_r^2\rangle $
and $E_\varphi = \frac{1}{2}\langle u_\varphi^2\rangle$
(here and in the following, $\langle \cdot \rangle$
denotes spatial average over the circular domain of radius $R$).

\begin{figure}[th!]
\includegraphics[width=0.21\textwidth]{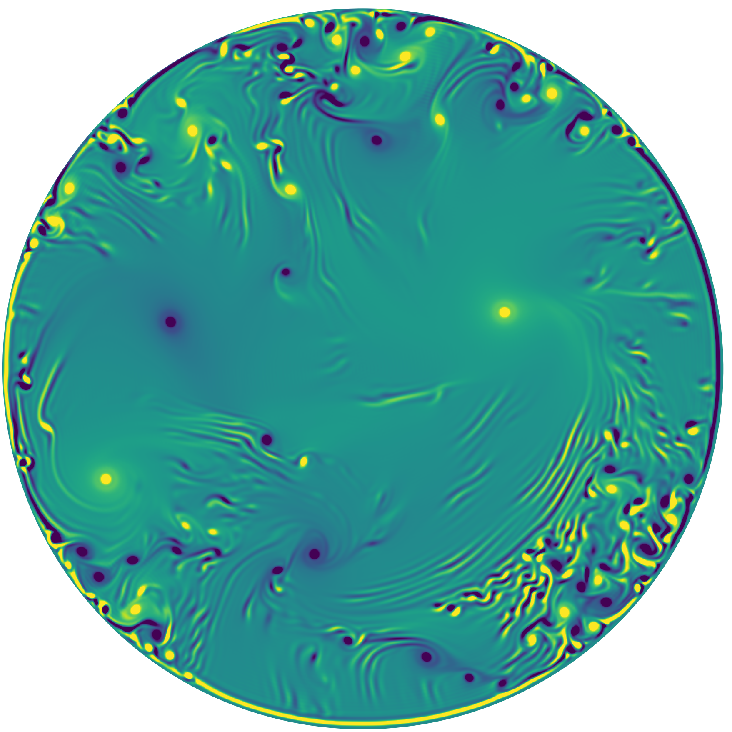} 
\includegraphics[width=0.21\textwidth]{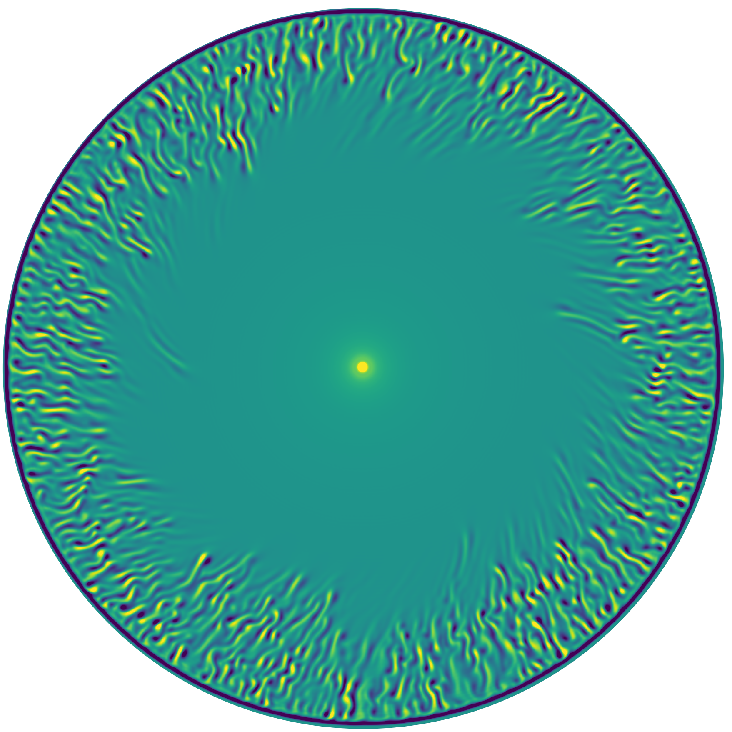} 
\includegraphics[width=0.0445\textwidth]{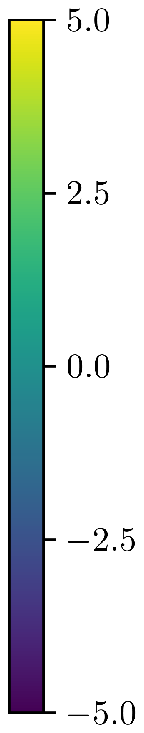} 
\caption{Vorticity field for 
the simulation with $R=31 \Lambda$ and $\alpha =-1.75$
at $t=210 \Lambda/U$ (left) and $t=550\Lambda/U$ (right). 
}
\label{fig1}
\end{figure}

We let the system evolve starting form a null velocity field
seeded with an infinitesimal random perturbation.
At the beginning of the simulation,
the swimmers organize in a large number of small-scale vortices, 
with equal probability of positive and negative vorticity
and homogeneous and isotropic spatial distribution. 
In this stage, the statistical properties of the flow are identical to
those observed in simulations with periodic boundary conditions
\cite{bratanov2015new,james2018vortex}. 
After a short time, the system evolves
towards an intermediate, turbulent-like regime,  
characterized by the presence of multiple large-scale vortices,
which move chaotically and are surrounded by regions of vorticity streaks
(see Fig.~\ref{fig1}, left panel).

\begin{figure}[th!]
\includegraphics[width=0.5\textwidth]{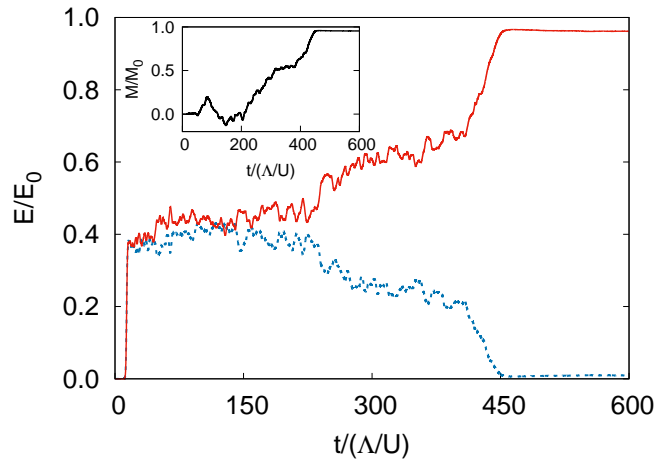}  
\caption{Temporal evolution of the radial and angular
components of the kinetic energy
$E_r$ (blue, dashed line),  
$E_\varphi$ (red, solid line) normalized with 
$E_0=\frac{1}{2} U^2$. 
The inset shows the evolution of the angular momentum
$M$ normalized with $M_0=\frac{2}{3} U R$. 
Simulation at $R=31 \Lambda$ and $\alpha =-1.75$.
}
\label{fig2}
\end{figure}

During this stage of the simulation we observe 
equipartition (with strong temporal fluctuations)
between the radial and angular components of the kinetic energy
(see Fig.~\ref{fig2}).
At later times, the system displays a rapid increase of $E_\varphi$
accompanied by the decrease of $E_r$,
which indicates the transition to a novel regime 
characterized by $E_\varphi \simeq E_0 \equiv \frac{1}{2} U^2$ 
and $E_r \simeq 0$.  
This corresponds to the self-organization of the swimmers
in a state of {\it circular flocking}, that is, 
a stationary, single, giant vortex
which spans the whole domain (see Fig.~\ref{fig1} right panel),
similar to that observed in experiments of bacterial suspension in a
viscoelastic fluid \cite{liu2021viscoelastic}.
Changing the initial condition of the flow, we
observed a strong variability of the transition times
from the intermediate turbulent regime to the giant-vortex state
(see Fig.2 in the Supplemental Material\cite{supplemental}). 

The formation of this large-scale structure causes a symmetry breaking 
of the angular momentum of the flow $M=\langle {\bm r} \times {\bm u} \rangle$. 
As shown in the inset of Fig.~\ref{fig2},
the values of $M$ fluctuate around zero
before the formation of the giant vortex.
Later, $M$ saturates to a constant value
$|M| \simeq M_0 \equiv \frac{2}{3} U R$
with definite sign. 

\begin{figure}[th!]
	\includegraphics[width=0.5\textwidth]{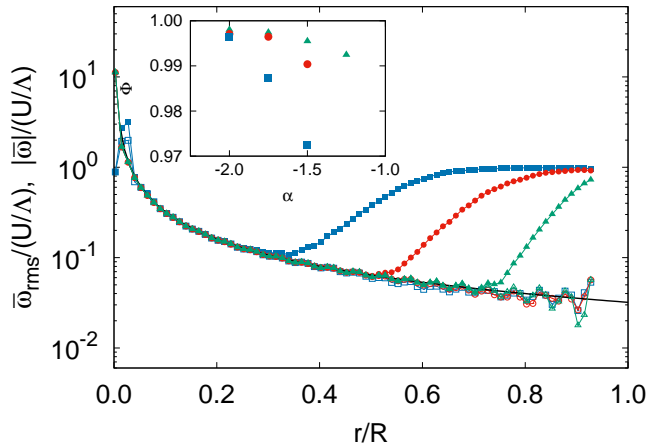} 
	\caption{
		Radial profiles of the vorticity, $\overline\omega(r)$ (empty symbols),
		and of the RMS vorticity,  $\overline\omega_{rms}(r)$ (filled symbols),
		for simulations with $R=31\Lambda$, 
                $\alpha=-1.5$ (blue squares),
		$\alpha=-1.75$ (red circles)
		and $\alpha=-2$ (green triangles). 
		The black line is the prediction
		$|\overline{\omega}(r)| = U/r$. 
		Inset: Mean value of the vortex order parameter $\Phi$
		as a function of $\alpha$ for
		$R=16\Lambda$ (green triangles),
		$R=23\Lambda$ (red circles),
		$R=31\Lambda$ (blue squares). 
	}
	\label{fig3}
\end{figure}

The time-averaged, mean radial vorticity profile of the giant vortex
$\overline{\omega}(r) = \frac{1}{2\pi r} \int \omega({\bm r}') \delta(|\bm{r}'|-r) d^2r'$
displays a power law behavior $\overline{\omega}(r) \propto 1/r$
in the region $\Lambda \lesssim r \lesssim R-\Lambda$
far from the boundaries and from the center (Figure~\ref{fig3}). 
A theoretical prediction for $\overline{\omega}(r)$
can be derived by assuming that the
radial component of the velocity vanishes, $u_r = 0$,
and that the angular component depends only on $r$
as $u_\varphi = r \Omega(r)$, where $\Omega(r)$ is the angular velocity. 
The resulting vorticity field is 
$\omega = {\bm \nabla} \times {\bm u} = 2\Omega(r) + r \partial_r \Omega(r)$.
Inserting these expressions in the equation for the vorticity,
which is obtained by taking the curl of Eq.~(\ref{eq:1}),
and imposing the stationarity condition,
one gets the following equation for $\Omega(r)$
\begin{equation}
  (\alpha + \Gamma_2 \nabla^2 +  \Gamma_4 \nabla^4)
  (2\Omega + r \partial_r \Omega)
  +\beta r^2 \Omega^2
  (4 \Omega + 3 r \partial_r \Omega) = 0\;. 
 \label{eq:2}
\end{equation}
With the further assumption 
(justified a posteriori
\footnote{The Swift-Hohenberg operator applied to a vorticity field $\overline{\omega}(r) = \pm U/r$ gives subleading terms of order 
$O((r/\Lambda)^{-3})$ which are negligible for $r \gg \Lambda$.}
)
that the Swift-Hohenberg term is negligible for $r \gg \Lambda$,  
Eq.~(\ref{eq:2}), admits the power-law solution
$\Omega(r) =  c r^\gamma$ with
$c=\pm \sqrt{-\alpha/\beta}$ and $\gamma=-1$.
This gives a prediction for the
radial profiles of
velocity $\overline{\bm u}(r) = \pm U \hat{\bm \varphi}$
and vorticity $\overline{\omega}(r) = \pm U/r$, 
which is in perfect agreement with our numerical findings
(see Figure~\ref{fig3}). 

The degree of order of the collective motion of the swimmers
can be quantified by the vortex order parameter
\cite{wioland2013confinement,lushi2014fluid,beppu2017geometry}
which is defined as 
$\Phi = (\langle |{\bm u} \cdot
\hat{\bm \varphi}|\rangle / \langle | {\bm u} | \rangle - 2/\pi)/(1-2/\pi)$. 
A velocity field oriented in the angular direction
${\bm u} \parallel \hat{\bm \varphi}$ gives $\Phi=1$,
while $\Phi=0$ corresponds to random-oriented velocity.
The values of $\Phi$ measured in the late stage are very close to $1$,
(see inset of Figure~\ref{fig3}), 
which indicates that the motion of the swimmers is highly ordered. 
The degree of order increases
reducing the radius $R$ of the domain and increasing $|\alpha|$.

Beside the giant vortex, Figure~\ref{fig1} also shows the presence
of vorticity fluctuations in an annular region close to the boundary.
These have the aspect of elongated structures,
slightly leaned in the direction of the mean flow of the vortex, 
which extend from the boundary toward the center of the domain. 
These structures are composed by pairs of vortical streaks with opposite sign,
corresponding to radial velocity jets 
with a typical transverse width of the order of $\Lambda$.
The average number of streaks in a domain of radius $R$
is therefore $N \simeq 2 R \sqrt{\Gamma_2/2\Gamma_4}$.
The formation of alternated streaks in the TTSH model
has been observed also in numerical simulations
in the absence of boundaries~\cite{mukherjee2021anomalous},
and it is responsible for superdiffusive behavior of Lagrangian 
tracers~\cite{singh2022lagrangian}. 

The intensity of the vorticity fluctuations can be quantified by the
RMS vorticity profile $\overline{\omega}_{rms}(r) = (\overline{\omega^2}(r))^{1/2}$
which is shown in Figure~\ref{fig3}.
Vorticity fluctuations are absent in the central region of the vortex
in which $\overline{\omega}_{rms}(r)$ coincides with
the mean radial profile $|\overline{\omega}(r)|$.
They appear at larger $r$, as shown by the increase of $\overline{\omega}_{rms}(r)$
which reaches an almost constant plateau close to the boundary
$\overline{\omega}_{rms}(r)\simeq U/\Lambda$.

\begin{figure}[th!]
\includegraphics[width=0.5\textwidth]{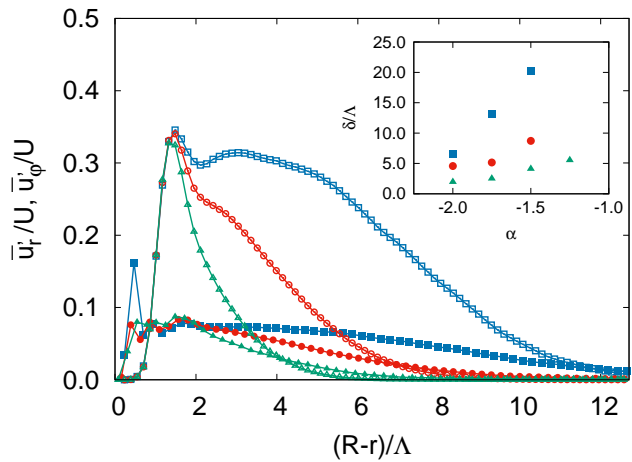} 
\caption{
Radial profiles of the radial and tangential components
of the velocity fluctuations,
$\overline{u'_r}(r)$ (empty symbols) and 
$\overline{u'_\varphi}((r)$ (filled symbols),
as a function of the distance from the boundary
for $\alpha=-1.75$, $R=31\Lambda$ (blue squares),
$\alpha=-1.5$, $R=23\Lambda$ (red circles)
and $\alpha=-1.25$, $R=16\Lambda$ (green triangles).
Inset: Width $\delta$ of the annular regions of the streaks 
as a function of $\alpha$ for
$R=16\Lambda$ (green triangles),
$R=23\Lambda$ (red circles),
$R=31\Lambda$ (blue squares). 
}
\label{fig4}
\end{figure}

Further details on the statistics of the streaks are revealed by the
profiles of radial and angular velocity fluctuations 
defined as $\overline{u'_r}(r) = (\overline{u_r^2}(r))^{1/2}$
and $\overline{u'_\varphi}(r) = (\overline{u_\varphi^2}(r)-\overline{u_\varphi}^2(r))^{1/2}$,  
shown in Figure~\ref{fig4}.
The radial component is predominant in the velocity field of the streaks. 
Close to the boundary, the ratio between the intensities of radial
and angular fluctuations is almost constant $\overline{u'_r}/\overline{u'_\varphi}\simeq 4.2$. 
The intensity of velocity fluctuations decays
at increasing the distance from the boundary $R-r$.

The width of the region in which the streaks are present
can be quantified as the distance $\delta$ from the boundary
at which the radial profile of the order parameter
exceeds a given threshold value $\overline{\Phi}(R-\delta) = \Phi_{thr}$.
The values of $\delta$ (with $\Phi_{thr}=0.9995$)
are reported in the inset of Figure~\ref{fig4}.  
We find that $\delta$ increases monotonically 
increasing the radius $R$ of the circular domain
and decreasing the parameter $|\alpha|$.
The scaling of $\delta$ as a function of the parameters 
of the model and of the radius $R$ remains an open question 
which deserves further theoretical studies. 

The formation of the giant vortex surrounded by streaks is the results of competing mechanisms
which can be understood by the comparison with the phenomenology observed in
numerical simulations with periodic boundary conditions.
In the latter case, the Landau potential and the self-propulsion term promote the development
of a flocking state, in which all the bacteria swim in the same direction 
with a constant speed \cite{toner1995long}. 
This collective ordered motion is destabilized by the Swift-Hohenberg operator
which causes the formation of vorticity streaks in the transverse direction
with respect to the mean flow~\cite{dunkel2013minimal}.  A possible explanation
of our findings is that the confinement in
circular domains drives the system toward a state of circular flocking
and then stabilizes it, preventing the formation of streaks in the center of the giant vortex.
Vorticity fluctuations are nonetheless generated close to the 
boundary by friction forces.
The vorticity production triggers the symmetry breaking of the angular momentum
and facilitates the formation of the giant vortex.

Despite this simple interpretation, the formation of the giant vortex,
is a highly non-trivial process which is far from being fully understood.
As shown in Figure~\ref{fig1}, the final state with a single vortex 
is achieved after a long turbulent regime in which several large-scale 
vortices compete with each other to prevail.  
We observed a strong variability of the duration of this intermediate regime
for different realizations of the flow, 
which confirms the complexity of this process. 
Moreover, at fixed $\alpha$ we found that there is a maximum size
of the domain which allows for the formation of the giant vortex.
For values of $R$ close to the maximum size, we observed the 
formation a giant vortex whose core consists of a binary rotating system 
of two small, equal-sign vortices 
(see Fig.1 in the Supplemental Material\cite{supplemental}). 
Increasing further the radius $R$ the evolution of the system remains 
in the turbulent regime characterized by multiple large-scale vortices 
which fail to merge in a single vortex during the simulation time. 

A comparison of the results of numerical simulations of the TTSH model
and experiments of bacterial turbulence in confined geometry
could shed new insight on this puzzling phenomenon.
A quantitative correspondence between our simulations and a feasible experimental setup
can be established by matching the parameters of the TTSH model 
with the typical values of the characteristic scale $\Lambda$ and velocity $U$
of the collective bacterial motion which are observed in experiments
(e.g., in~\cite{sokolov2012physical,wensink2012meso,dunkel2013fluid}).
As an example, by fixing $\Lambda \simeq 25 \mu m$ and $U \simeq 50 \mu m /s$
the values of the radius $R$ of the circular domain considered in our study
correspond in physical units to the range $R \simeq (400 -800) \mu m$,
the values of the parameter $\alpha$ are in the range 
$- \alpha \simeq (1.4 - 1.8)s^{-1}$  
and the typical time required to observe the formation of the giant vortex 
is of the order of minutes (see Figure~\ref{fig2}).
These spatial and temporal scales are easily accessible in experiments
of dense bacterial suspensions, such as those of {\it Bacillus subtilis}.

\section*{Acknowledgments}
We acknowledges support from the Departments of Excellence grant (MIUR)
and INFN22-FieldTurb.
We thank M. Cencini for useful comments and suggestions. 


\bibliography{biblio}

\end{document}